\documentclass [english,aps,prb,twocolumn,amsmath,amssymb,showpacs,showkeys]{revtex4}
\usepackage[T1]{fontenc}
\usepackage[latin1]{inputenc}
\usepackage{graphicx}
\usepackage{gensymb}
\makeatletter
\usepackage{epsfig}
\usepackage{prettyref}
\usepackage{babel}
\usepackage{subfigure}
\usepackage{color}

\makeatother
\begin{document}

\title{Spin glass behavior of gelatin coated NiO nanoparticles}
\author{Vijay Bisht}
\email{vkbisht81@gmail.com}
\author{K.P.Rajeev}
\email{kpraj@iitk.ac.in} \affiliation {Department of Physics,
Indian Institute of Technology Kanpur 208016, India}
\author{Sangam Banerjee}
\affiliation {Surface Physics Division, Saha Institute of Nuclear
Physics Kolkata, 700 064, India}

\begin{abstract}

We report magnetic studies on gelatin coated NiO nanoparticles of
average size 7~nm. Temperature and time dependence of \emph{dc}
magnetization, wait time dependence of magnetic relaxation
(aging), memory effects in the \emph{dc} magnetization and
frequency dependence of \emph{ac} susceptibility have been
investigated. We observe that the magnetic behavior of coated NiO
nanoparticles differs substantially from that of bare
nanoparticles. The magnetic moment of the coated particles is
highly enhanced and the ZFC magnetization data displays a sharp
peak ($T_{\textnormal{p1}} \approx 15$~K) at a low temperature in
addition to a usual high temperature peak ($T_{\textnormal{p2}}
\approx 170$~K). We observe that this system exhibits various
features characteristic of spin glass like behavior and
$T_{\textnormal{p2}}$ corresponds to the average freezing
temperature. We argue that this behavior is due to surface spin
freezing within a particle. The nature of the low temperature peak
is however ambiguous, as below $T_{\textnormal{p1}}$ some features
observed are characteristic of superparamagnetic blocking while
some other features correspond to spin glass like behavior.

  \end{abstract}
 \maketitle

\section{INTRODUCTION}
\label{INTRODUCTION}
 There has been a renewed interest in magnetic
nanoparticles in the past several decades because of the promise
of various possible technological applications they hold as well
as from the perspective of fundamental
understanding.\cite{Steen,Dormann} Below a certain size, a
ferromagnetic particle consists of a single domain and behaves as
a giant magnetic moment. N\'{e}el proposed that just as a
ferromagnetic particle, a tiny antiferromagnetic particle can also
develop a net magnetic moment due to uncompensated spins at its
surface. \cite{Neel} The behavior of an ensemble of
 non interacting particle moments is expected to be
superparamagnetic.\cite{Dormann} On the other hand if the
particles interact with each other they can give rise to superspin
glass behavior.\cite{Bedanta} Further, surface effects are also
important in nanoparticles because of their large
surface-to-volume ratio. For instance it has been shown recently
that spin glass behavior can arise within an individual
nanoparticle due to the freezing of spins at its
surface.\cite{Tiwari,Winkler,RHKodama}

 Nickel oxide (NiO) is an antiferromagnetic  material with N\'{e}el
temperature ($T_{\textnormal{N}}$) 523~K. There have been a large
number of studies on NiO nanoparticles and indeed it is the most
well studied among the antiferromagnetic nanoparticles. These
studies include temperature dependence of \emph{dc} and \emph{ac}
susceptibility, their field and frequency dependence, hysteresis
and exchange bias, time dependence of magnetization and related
dynamic
effects.\cite{Bisht,Tiwari,Seehra,Winkler,Richardson,Makhlouf,
Jagodic,Yuko,Ghosh,Park, Bodker,Shim,Thota,Mseehra,Mandal} Certain
features shown by NiO nanoparticles are characteristic of both
superparamagnetism and spin glass behavior viz. a bifurcation in
FC and ZFC magnetization, a peak in ZFC magnetization and slow
magnetic relaxation. As a result there have been claims of
superparamagnetism as well as spin glass behavior in this system.
\cite{Ghosh,Jongnam,Richardsona,Fatemeh,Myge,Mseehra,Yuko,Tiwari,
Winkler,Thota,Mandal} However, it has been shown that the
temperature dependence of magnetization of this system above the
bifurcation temperature cannot be described by the Langevin
function or a modified version of the same tailored for
antiferromagnetic particles. We note that this is contrary to what
is expected of a superparamagnetic system.\cite{Makhlouf}

Below the bifurcation temperature,  spin glass or spin glass like
features have been observed in this system.\cite{Tiwari,
Winkler,Thota,Bisht,Mandal} In a recent paper, we reported aging
and memory effects in bare NiO nanoparticles and the results show
that this system indeed shows spin glass like
behavior.\cite{Bisht} Tiwari et al. have argued that such behavior
arises due to freezing of surface spins on individual particles as
the interactions between the particles are too weak to give rise
to the observed large freezing temperatures.\cite{Tiwari} The
origin of spin glass behavior in NiO nanoparticles is still
somewhat controversial as there is no universal agreement on
whether it is due to interparticle interactions or due to surface
spin frustration within individual particles. A possible way to
settle this issue would be through a study of the magnetic
behavior of non-interacting particles. We can reduce the
inter-particle interactions by increasing the distance between
them. One way of doing this would be to coat the particles or
disperse them in some medium. There have been some works on coated
and dispersed NiO nanoparticles and nanorods and widely varying
results have been
reported.\cite{Winkler,Ghosh,Shim,Hshim,Seehra,Bodker,Myge} In
these works, the upper broad peak is generally associated with
superparamagnetic blocking. Further it has been observed that the
coating tends to decrease the interactions which manifests as a
lowering of the blocking temperature. Another sharp peak at a much
lower temperature is seen in some reports and the origin of this
peak has not been accounted for in most of the
works.\cite{Seehra,Bodker,Myge,Hshim,Shim} However some authors
associate this peak to surface spin freezing.\cite{Winkler,Thota}
To further complicate matters, some authors observe only the low
temperature peak with the upper broad peak missing in coated
particles and in one report this peak has been seen even in a bulk
sample.\cite{Ghosh,Park,Jagodic} We, therefore, felt that it will
be worth our while to carry out a systematic study on the magnetic
behavior of coated NiO nanoparticles to clear the air.

\begin{figure}[t]
\begin{centering}
\includegraphics[width=1\columnwidth]{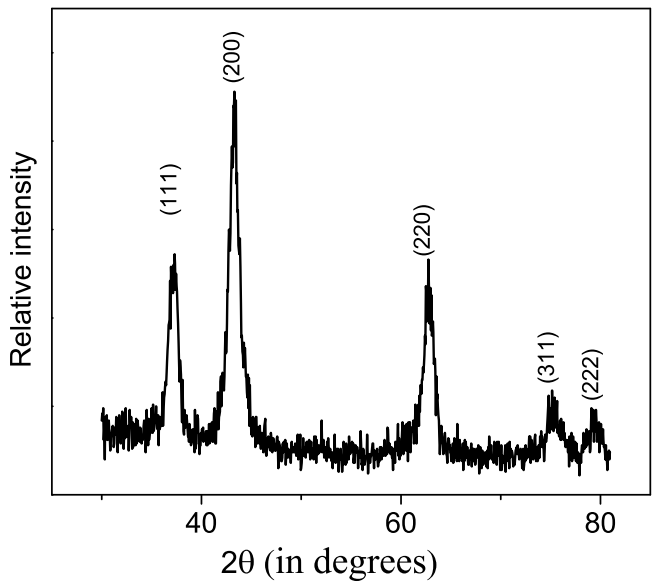}
\par\end{centering}
\caption{XRD pattern of the sample heated at 350\degree C for 15
hours. All the peaks correspond to those of NiO. } \label{fig:XRD}
\end{figure}
\begin{figure}[b]
\begin{centering}
\includegraphics[width=1\columnwidth]{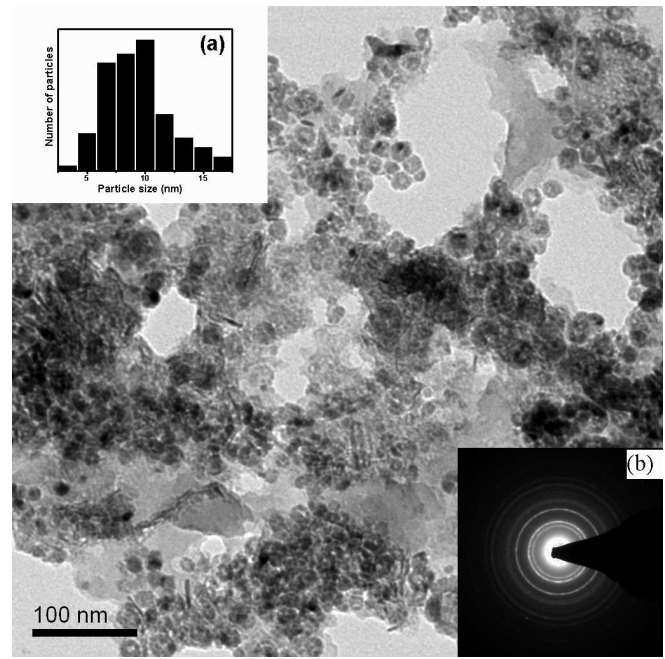}
\par\end{centering}
\caption{TEM image of the sample. Inset: (a) Histogram of the
particle size distribution. Total Number of particles considered
is 110. (b) Selected area diffraction (SAD) pattern. }
\label{fig:TEM}
\end{figure}

\section{EXPERIMENTAL DETAILS}
\label{EXPERIMENTAL DETAILS}
 Gelatin coated NiO nanoparticles are
prepared by a sol gel method as described in detail by Meneses et
al.\cite{Meneses} In brief, 2.5 g of gelatin was dissolved in 100
ml distilled water while stirring continuously  at 60\degree C.
100 ml of aqueous solution of 2.5 g of NiCl$_2$.6H2O (99.99\%) was
added at the same temperature to the above solution. An aqueous
solution of NaOH was added to this mixture till the pH  became 12.
This mixture was then cooled at room temperature to form a gel,
which was  heated at 80 \degree C for 36 hours to obtain a
precursor. Nickel oxide nanoparticles were prepared by heating
this precursor at 350\degree C for 24 hours . The sample was
characterized by X-ray diffraction (XRD) using a Seifert
diffractometer with Cu~K$\alpha$ radiation and Transmission
electron microscopy (TEM) using  FEI Technai 20 U Twin
Transmission Electron Microscope. The percentage of gelatin by
mass was estimated using thermo-gravimetric analysis(TGA) to be
42\% and the average thickness of gelatin shell estimated turned
out to be 5~nm. All the magnetic measurements were done with a
SQUID magnetometer (Quantum Design, MPMS XL5).

\section{RESULTS AND DISCUSSION}
\label{RESULTS AND DISCUSSION}
\subsection{Particle size}
\label{Particle size}
 The XRD pattern of the sample shown in
Figure \ref{fig:XRD} corresponds to that of pure NiO which has FCC
structure. The average particle size was estimated to be 7~nm from
the width of XRD peaks (111), (200) and (220) using the Scherrer
formula. TEM image of the sample is shown in Figure \ref{fig:TEM}
and the insets (a) and (b) of this  figure show the particle size
distribution and the selected area diffraction (SAD) pattern. It
can be seen that the particles are more or less of spherical shape
and  the particle size was estimated to be 9.3 nm with a standard
deviation 2.8 nm. The SAD pattern consists of concentric
diffraction rings with different radii. The diameter of a
diffraction ring in the SAD pattern is proportional to
\(\sqrt{h^{2} + k^{2} + l^{2}}\), where \(({h}{k}{l})\) are the
Miller indices of the planes corresponding to the ring. Counting
the rings from the center 1st, 2nd, 3rd, 4th and 5th rings
correspond to (111), (200), (220), (311) and (222) planes
respectively, as would be expected in the case of a material with
FCC crystal structure.

\subsection{Temperature and field dependence of magnetization}
\label{Temperature and field dependence of magnetization}
\begin{figure}[t]
\begin{centering}
\includegraphics[width=1\columnwidth]{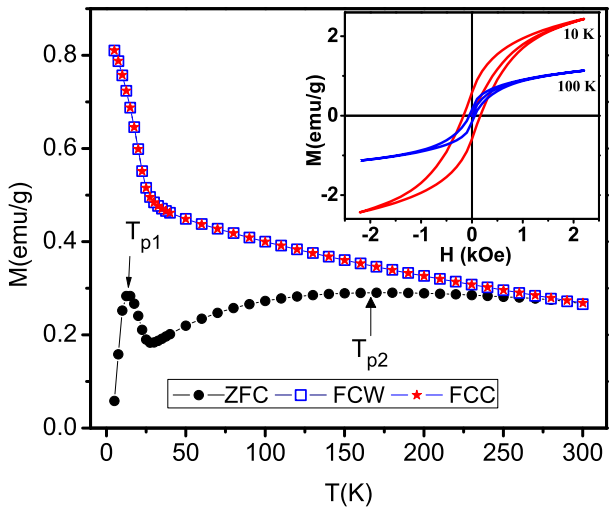}
\par\end{centering}
\caption{(Color online) FCC, FCW and ZFC magnetization data at a
magnetic field of 100~Oe. Inset shows the hysteresis loops at 10~K
and 100~K including the virgin curve.} \label{fig:FCZFC}
\end{figure}

 The temperature dependence of magnetization is done under field cooled
(FC) and zero field cooled (ZFC) protocols at 100~Oe. See Figure
\ref{fig:FCZFC}. It can be seen that the FC and ZFC magnetization
curves bifurcate slightly below 300~K and the magnetization data
taken during heating (FCW) and that taken while cooling (FCC) in
the FC protocol are essentially the same. Further it can be
observed that there are two peaks in the
 ZFC magnetization; the first $(T_{\textnormal{p1}})$ is a sharp one
at about 14~K and the second $(T_{\textnormal{p2}})$ is a broad
one around 170~K. The FC magnetization shows a steep low
temperature rise starting at about 30~K and keeps on rising till
the lowest temperature of measurement, a characteristic feature
seen in superparamagnets.\cite{Sasaki}

 In bare nanoparticles, usually a single broad peak, corresponding to
$T_{\textnormal{p2}}$, is observed in the magnetization vs.
temperature plot.\cite{Richardson,Bisht,Tiwari,Makhlouf} However,
there are some reports on bare, dispersed and coated nanoparticles
where two peaks have been
observed.\cite{Winkler,Thota,Hshim,Seehra,Jagodic} Some workers
even report a single peak at low temperature in coated
nanoparticles with the upper broad peak missing\cite{Ghosh} and to
confound matters even further, some authors have observed
$T_{\textnormal{p1}}$ even in bulk samples.\cite{Jagodic} Winkler
et al. have observed two peaks in ZFC magnetization for NiO
nanoparticles of size 3~nm; occurring at 17~K and 70~K for bare
particles and at 15~K and 60~K for dispersed
particles.\cite{Winkler} They found that the high temperature peak
in the \emph{ac} susceptibility data follows the Arrhenius law
like superparamagnets while the lower peak follows a power law
similar to spin glasses. Further the shape of virgin curve in the
hysteresis loop below the low temperature peak is S-shaped, a
feature seen in canonical spin glasses while well above the lower
peak temperature, this feature is absent. Thus they associate the
upper peak with superparamagnetic blocking of core moments and the
lower peak to surface spin glass freezing. However Tiwari et al.
have reported a single broad peak in the ZFC magnetization as well
as in \emph{ac} susceptibility in 5~nm bare nanoparticles at about
150~K and they have shown that the system shows spin glass
features. For instance, the  value of relative shift of \emph{ac}
susceptibility peak per decade of frequency lies in a  range
expected for spin glasses, field dependence of peak temperature
follows Almeida-Thouless (AT) line and the high field data obeys
\emph{dc} scaling law for spin glasses.\cite{Tiwari}

We carried out hysteresis measurements in the field range
$-2.2$~kOe to $+2.2$~kOe at temperatures 10~K and 100~K. The
results are shown in the inset of Figure \ref {fig:FCZFC} and it
can be seen that the system shows hysteresis at both 10~K and
100~K with a larger coercivity at 10~K. In contrast to what
Winkler et al. got, the virgin curve in this case is not S-shaped
either at 10~K or at 100~K.\cite{Winkler} An S-shaped  virgin
curve is a feature observed in canonical spin glasses, but we do
not see it in our system. \cite{Mydosh}

\begin{figure}[!t]
\begin{centering}
\includegraphics[width=1\linewidth]{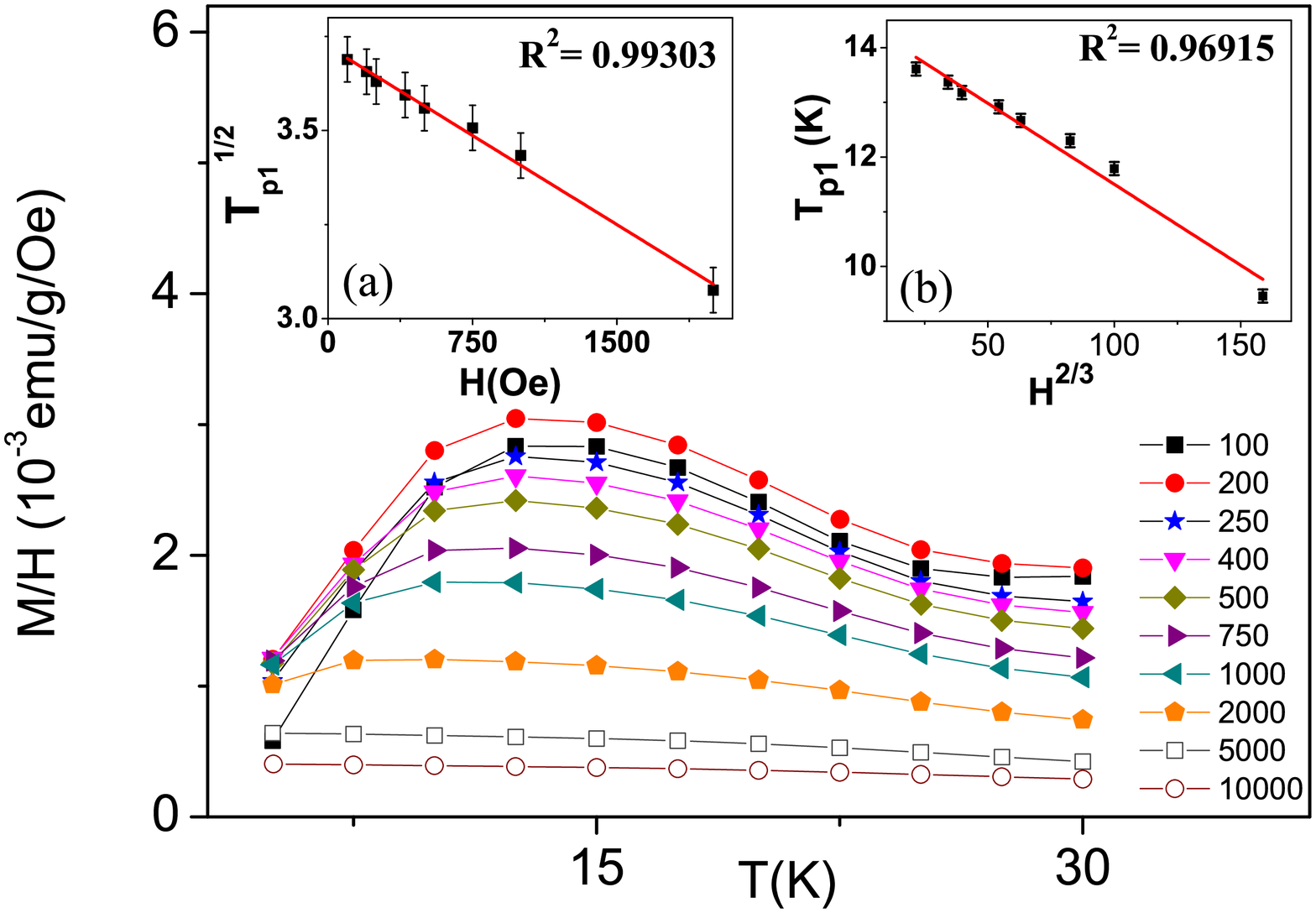}
\par\end{centering}
\caption{(Color online) Field dependence of ZFC magnetization data
 for various fields at low temperatures in the vicinity of
  $T_{\textnormal{p1}}$. Insets show plots of
  (a)  $T_{\textnormal{p1}}^\frac{1}{2}$  vs $H$ and
  (b) $T_{\textnormal{p1}}$ vs $H^\frac{2}{3}$.}
  \label{fig:lower peak}
\end{figure}

\begin{figure}[!b]
\begin{centering}
\includegraphics[width=1\linewidth]{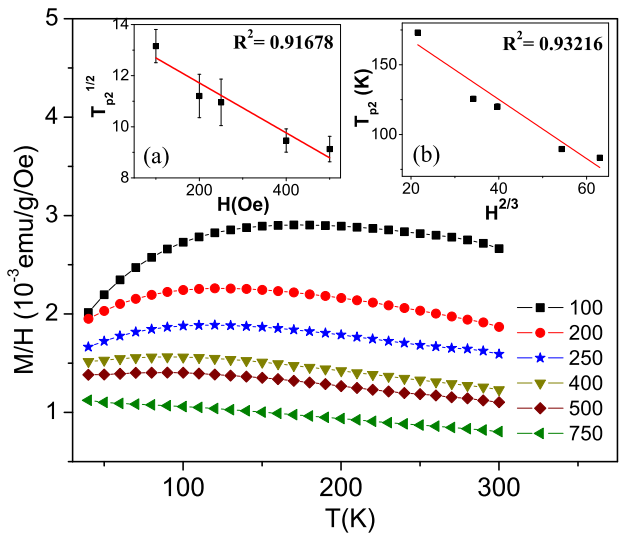}
\par\end{centering}
\caption{(Color online) Field dependence of ZFC magnetization data
 for various fields at higher temperatures in the vicinity of
  $T_{\textnormal{p2}}$.  Insets show plots of
  (a) $T_{\textnormal{p2}}^\frac{1}{2}$  vs $H$ and (b)
T$_{\textnormal{p2}}$ vs $H^\frac{2}{3}$.}
\label{fig:higher peak}
\end{figure}

To investigate the field dependence of ZFC magnetization, we
carried out experiments at various fields in the field range
100~Oe to 10~kOe. These data are shown in Figures \ref{fig:lower
peak} and
 \ref{fig:higher peak}. It can be observed that both the
peaks $(T_{\textnormal{p1}}$ and $T_{\textnormal{p2}})$ shift to
lower temperatures with increasing field; the dependence being
weaker for the lower peak. $T_{\textnormal{p2}}$ disappears above
an applied field of 750 Oe while $T_{\textnormal{p1}}$ disappears
only above 2~kOe. For superparamagnets  the field dependence of
peak temperature, $T_{\textnormal{p}}$, is expected to be given by
\cite{Bitoh,RKZheng}
 \begin{equation}
\label{Super-para-blocking} T_{\textnormal{p}} \propto V
\left(1-\frac{H}{H_{\textnormal{K}}}\right)^{2}, \; \; \;
  0 \leq H \leq  H_{\textnormal{K}}
\end{equation} 
where $V$ is the volume of a particle and $H_{\textnormal{K}}$ is
a positive constant depending on anisotropy of the system.  Coming
to the case of spin glasses we note that the stability limit of
spin glasses is defined by the AT line in the $H-T$ phase diagram,
below which the spin glass state is stable.\cite{Almeida,Mydosh}
Indeed in many spin glass systems, the field dependence of peak
temperature is known to follow the AT line given by the
equation:\cite{ATline,Tiwari}

\begin{equation}
\label{Almeida-Thouless} H \propto \left(1 -
\frac{T_{\textnormal{p}}}{T_{\textnormal{f}}}\right)^\frac{3}{2},
 \; \; \; 0\leq
{T_{\textnormal{p}}}\leq{T_{\textnormal{f}}}
 \end{equation}
where $T_{f}$ is the spin glass transition temperature in zero
applied field. Thus, in a superparamagnetic system
$T_{\textnormal{p}}^\frac{1}{2}$ should be linearly related to $H$
whereas in spin glasses $T_{\textnormal{p}}$ should decrease
linearly with $H^\frac{2}{3}$. In the insets of Figures \ref{fig:lower
peak} and \ref{fig:higher peak}, we show the plots of
T$_{\textnormal{p}}$ vs $H^\frac{2}{3}$ and $T_{\textnormal{p}}^\frac{1}{2}$
vs $H$  for both peaks. The goodness of the fits can be judged  by
the coefficient of determination ($R^2$) which are shown in the
corresponding plots. It can be seen that for the lower peak the
 superparamagnetic fit is better while for the upper peak, the AT line
fit is better. Thus from these experiments, one can hazard the
guess that the lower peak arises due to superparamagnetic blocking
while the upper one corresponds  to spin glass behavior.

\begin{figure}[!t]
\begin{centering}
\includegraphics[width=1\linewidth]{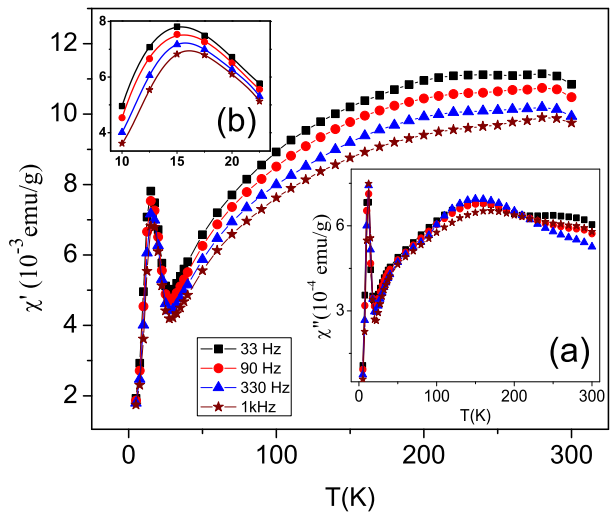}
\par\end{centering}
\caption{(Color online) Temperature dependence of the real part of
\emph{ac} susceptibility for various frequencies with an ac field
of 3~Oe. Insets: (a) Temperature dependence of the imaginary part
of the susceptibility.  (b) A magnified view of low temperature
peak of real part shown in the main panel. Lines have been drawn
to guide the eyes. }
\label{fig:acsusceptibility}
\end{figure}

We note that the broad peak, $T_{\textnormal{p2}}$, in coated
nanoparticles appears at about 170~K at low field (Figure
\ref{fig:FCZFC}) which is quite close to the corresponding value
150~K seen in bare particles of comparable size as reported by
Tiwari et al. \cite{Tiwari} Apparently interparticle interactions
have little influence on $T_{\textnormal{p2}}$ and thus cannot be
contributing to spin glass behavior. Indeed Tiwari et al. estimate
that the dipolar interactions between bare particles can give rise
to freezing temperatures at most a few Kelvins. \cite{Tiwari} Thus
we conclude that in NiO nanoparticles, whether bare or coated, the
interparticle interactions are quite small and cannot give rise to
spin glass like behavior.

\subsection{Particle moment}
\label{Particle moment}
 N\'{e}el proposed that small particles of
antiferromagnetic materials can possess a net magnetic moment due
to incomplete compensation of spins between atoms on two
sublattices.\cite{Richardsona} The number of uncompensated spins,
$p$, is roughly proportional to $n^x$ where $n$ is the total
number of atoms in the particle and $x$ can be $1/3$, $1/2$ or
$2/3$ depending on the shape of the particle and the arrangement
of atoms in it.\cite{Richardsona} The particle moment depends on
$p$ and thus on the particle size. Kodama et al. have estimated a
magnetic moment $80\mu_{\textnormal{B}}$ per particle  for bare
15~nm NiO particles using N\'{e}el's two sublattice model and
found that this value is too small compared to
$700\mu_{\textnormal{B}}$, a value estimated from experimental
data.\cite{Kodama} They proposed the existence of a
multi-sublattice ordering in NiO nanoparticles to account for the
anomalously high magnetic moment.

A linear extrapolation of the high field magnetization data at
10~K (shown in the inset of  Figure \ref{fig:FCZFC}) gives an
estimated moment of about $1350~\mu_{\textnormal{B}}$ for the
gelatin coated 9~nm particles. We see  that the particle magnetic
moment  increases roughly by 2.5 times  on coating bare
nanoparticles with gelatin.\cite{moment}  A similar enhancement in
the particle moment was observed by Winkler et al. on dispersing
3~nm particles in a non magnetic matrix.\cite{Winkler} This
observation is puzzling because it is unclear how a non-magnetic
coating leads to an enhancement of magnetic moment of a particle.
Winkler et al. have argued that the absence of demagnetizing
character of interparticle interactions is responsible for the
increase in magnetization in coated nanoparticles. We disagree
with this argument because as we had discussed earlier (section
\ref{Temperature and field dependence of magnetization}), the
interparticle interactions are quite small and hence cannot give
rise to such an enormous decrease in the magnetic moment of bare
particles as opposed to coated particles.

\begin{figure}[!b]
\begin{centering}
\includegraphics[width=1\linewidth]{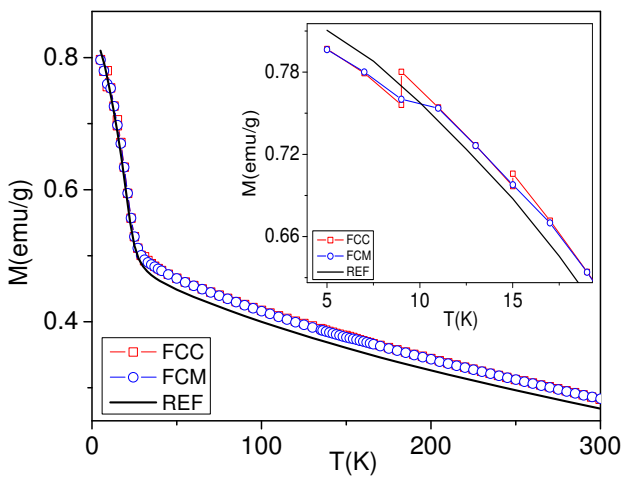}
\par\end{centering}
\caption{(Color online) Memory experiments in FC protocol with
stops of one hour taken at temperatures 8~K, 15~K and 150~K.
Magnetic field was switched off during the stops and switched on
before resuming further cooling.}
\label{fig:FCM}
\end{figure}

\subsection{\emph{ac} susceptibility}
\label{ac susceptibility}
 We measured the temperature dependence
of ac susceptibility at several frequencies: 33, 90, 330 and
1000~Hz. The sample is cooled from room temperature to 5~K in a
zero magnetic field and a probing \emph{ac} magnetic field of
amplitude 3~G is applied to measure the susceptibility as the
temperature is increased to 300~K. In Figure \ref
{fig:acsusceptibility}, the real part, $\chi'$, of the \emph{ac}
susceptibility is shown and the inset (a) displays the imaginary
part, $\chi''$. Inset (b) shows a magnified view of the low
temperature peak.  We note that the real part ($\chi'$) has a
sharp peak near 16~K and a broad high temperature peak between
200~K and 300~K. This broad high temperature peak can be observed
more clearly in the imaginary part. As the frequency is raised the
value of $\chi'$ decreases and the peaks shift slightly to higher
temperatures. A quantitative measure of the variation of peak
temperature with frequency is the relative shift in peak
temperature, ${\Delta T_{\textnormal{p}}}/{T_{\textnormal{p}}}$,
per decade of frequency. This quantity lies between 0.0045 and
0.06 for many canonical spin glasses. \cite {Mydosh}  For
ferritin, a known superparamagnet, its  value is approximately
$0.13$ and for another superparamagnet a-(Ho$_2$O$_3$)(B$_2$O$_3$)
it is $0.28$. \cite {Mydosh,Tiwari} In the present case, for the
lower peak, using the real part of susceptibility, this value
comes out to be $0.046$ and using the imaginary part, its value is
$0.067$. For the upper peak, using the imaginary part, this value
turns out to be $0.0553$. Thus for both  the peaks, the relative
shift lies in the range observed in spin glass and spin glass like
systems and provides an empirical evidence in support of spin
glass like behavior.

\begin{figure}[!t]
\begin{centering}
\includegraphics[width=1\linewidth]{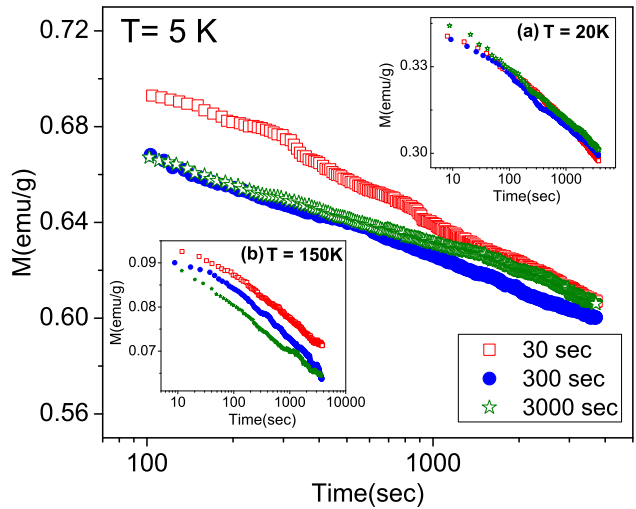}
\par\end{centering}
\caption{(Color online) Aging experiments in FC protocol  at
temperatures 5~K, 20~K and 150~K with wait times
$t_\textnormal{W}$ = 30, 300 and 3000 seconds.}
\label{fig:FCaging}
\end{figure}

\begin{figure}[!b]
\begin{centering}
\includegraphics[width=1\linewidth]{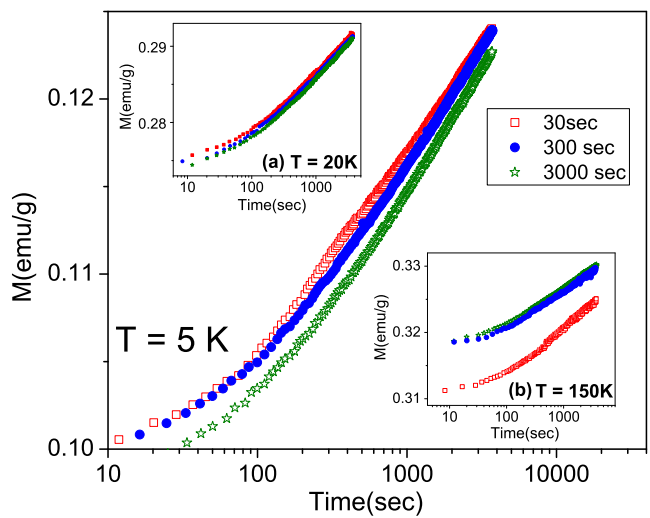}
\par\end{centering}
\caption{(Color online) Aging experiments in ZFC protocol  at
temperatures 5~K, 20~K and 150~K with wait times
$t_\textnormal{W}$ = 30, 300 and 3000 seconds.}
\label{fig:ZFCaging}
\end{figure}

\subsection{Memory and Aging Experiments}
\label{Memory and Aging Experiments}
 In the past several years, aging and memory effects have
been investigated in many nanoparticle systems using \emph{ac}
susceptibility and low field \emph{dc} magnetization measurements
with various temperature and field protocols.\cite{Bedanta,Sasaki}
 It has been seen that  superparamagnets as well as spin glasses
 show these effects in  FC protocol. However only spin glasses
show aging and memory in ZFC protocol.\cite{Sasaki}  We have
reported memory effects in bare NiO nanoparticles in both FC and
ZFC protocols in a previous work.\cite{Bisht} Therefore, it will be
interesting to investigate these effects in coated nanoparticles
where interactions between the particles should be negligible.

We carried out memory experiments in both FC and ZFC protocols
with stops of one hour taken at 8~K, 15~K and 150~K. The procedure
of these experiments is as follows. In the FC protocol, the system
is cooled in the presence of a magnetic field (100~Oe) to 5 K with
intermittent stops of one hour at 8~K, 15~K and 150~K, with the
field switched off during the stops. The magnetization is measured
while cooling and then during subsequent heating. The results of
memory experiments in FC protocol are shown in Figure \ref
{fig:FCM}. We found that at 150~K, there are no indications of
memory but at 8~K and 15~K, memory is present as is evident from a
small jump in the magnetization at these temperatures.
 However, these effects are much weaker than those
observed in bare NiO nanoparticles.\cite{Bisht} In the ZFC
protocol, to begin with we record the ZFC magnetization data
normally and then with stops of one hour at 8~K, 15~K and 150~K
while cooling. We observed that there is no significant difference
between these two data and this shows that the system has no ZFC
memory.

These experiments indicate that the lower peak can correspond to
superparamagnetic blocking as the memory is present only in FC
measurements and not in ZFC measurements. On the other hand
absence of memory at 150~K in both FC and ZFC protocols is quite
unusual as it does not correspond to either spin glass like or
superparamagnetic behavior. However this apparent absence of
memory could be because of some transition associated with the lower peak $T_{\textnormal{p1}}$
which can wipe out the memory of stops taken at higher
temperatures.

 In addition to memory effects, aging has been used as a tool to
distinguish superparamagnets and spin glasses. We recall that
aging is seen in FC protocol in both superparamagnets and spin
glasses while in ZFC protocol it is seen only in spin glasses.
\cite{Bedanta,Sasaki} We investigated aging at three different
temperatures 5~K, 20~K and 150~K in both FC and ZFC  protocols.
To check for FC aging the sample is cooled in a field of 100 Oe
to the temperature of interest, and after waiting for a specified
duration (wait time, $t_\textnormal{W}$)  the field is switched
off. Subsequently  the magnetization data is recorded as a
function of time. These data are presented in Figure \ref
{fig:FCaging}. Similarly, in the corresponding ZFC aging
experiment, the sample is cooled in a zero field   to the
temperature of interest, and after waiting for a specified wait
time  the field (100 Oe) is switched on; magnetization as a
function of time is recorded subsequently. The ZFC aging data are
presented in Figure \ref{fig:ZFCaging}. We note that a good amount
of aging is discernible in both FC  and ZFC magnetizations at 5~K
and 150~K but not at 20~K. The presence of aging in ZFC protocol
at 5~K and 150~K is a strong evidence which  supports the thesis
that the system is  spin glass like.

\subsection{Static Scaling}
\label{Static Scaling}
 Static critical scaling has been widely
used as an evidence for phase transition in spin glasses and an
appropriate quantity to examine the critical behavior is the
nonlinear susceptibility, $\chi_{\textnormal{NL}}$, given
as\cite{Mydosh}
\begin{equation}
\chi_{\textnormal{NL}} = \chi_{\textnormal{0}} - \frac{M}{H} =
-(\chi_{\textnormal{2}} H^{2} + \chi_{\textnormal{4}} H^{4} +
...).
\end{equation}
It should be noted that $\chi_{\textnormal{NL}}$ should diverge in
the critical region as $\chi_{\textnormal{2}}$,
$\chi_{\textnormal{4}}$, ... are divergent in that region. To
describe $\chi_{\textnormal{NL}}$ in the critical region, the
following scaling equation has been proposed \cite {Geschwind}

\begin{equation}
\chi_{\textnormal{NL}} \propto H^{2\beta/(\beta + \gamma)} \bar{G}
(t/H^{2/(\beta + \gamma)}), \label{eq:scaling}
\end{equation}

where $t$ is the reduced temperature
$(\frac{T-T_{\textnormal{f}}}{T_{\textnormal{f}}})$, $\beta$ and
$\gamma$ are critical exponents of the spin glass order parameter
and $\bar{G}$ is the scaling function.

\begin{figure}[t]
\begin{centering}
\includegraphics[width=1\linewidth]{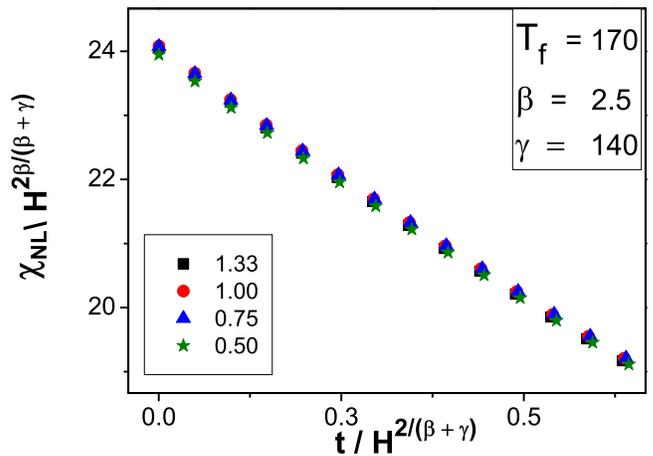}
\par\end{centering}
\caption{(Color online) Linear scaling plot of the \emph{dc}
nonlinear susceptibility. The scaled curve is obtained using
$T_{\textnormal{f}} = 170~K$, $\gamma = 140$ and $\beta = 2.5$.
($H$ is in units of Oe and $\chi_{\textnormal{NL}}$ is in units of
emu/gOe.) }
\label{fig:scaling}
\end{figure}

To demonstrate scaling, the parameters $\beta$, $\gamma$ and
$T_{\textnormal{f}}$ are selected so that all the data points
taken at various fields are judged to fall on a single master
curve on a plot of $\chi_{\textnormal{NL}}/H^{2\beta/(\beta +
\gamma)}$ vs. $t/H^{2\beta + \gamma}$. Figure \ref {fig:scaling}
shows the scaling plot of our data using Equation (\ref
{eq:scaling}). It is clear that four
 data sets taken at different magnetic fields are falling
 well on a master curve. It has been seen that several sets of
$\beta$, $\gamma$ and $T_{\textnormal{f}}$ can give reasonably
good plots.\cite {Geschwind} In Figure \ref {fig:scaling}, we used
$T_{\textnormal{f}} = 170~K$, $\gamma = 140$ and $\beta = 2.5$.
The values of critical exponents  $\beta$ and $\gamma$ should be
unity according to mean field theory. However it has been seen
that the values determined from experiments can be much larger
than unity.\cite {Fischer} Thus the magnetization data follows
scaling laws confirming spin glass behavior in this system. It can
be however noted that the non linear susceptibility does not
diverge in contrast to canonical spin glasses and the reasons for
this could  be the finite size of the system  and the distribution
of freezing temperatures due to particle size distribution as has
been discussed by Tiwari et al.\cite{Tiwari}

 \section{Conclusion}
 \label{Conclusion}
 We find that the behavior of gelatin coated NiO nanoparticles is
more intriguing than that of bare particles. The particle magnetic
moment is enhanced several times on coating and the reasons for
this phenomenon are not clear. An additional peak
($T_{\textnormal{p1}}$) is observed in the ZFC magnetization data
at 14~K which is usually not present in bare nanoparticles. The
field dependence of $T_{\textnormal{p2}}$ in ZFC magnetization
follows the AT line as one would expect in the case of a spin
glass. Further, ${\Delta
T_{\textnormal{p}}}/{T_{\textnormal{p}}}$, per decade of frequency
in \emph{ac} susceptibility  lies in the range seen in spin
glasses.  Strong aging effects have been observed at 150~K in both
FC and ZFC protocols, again a feature characteristic of spin glass
like systems. The \emph{dc} scaling analysis presents conclusive
evidence in support of spin glass behavior with
$T_{\textnormal{f}} = 170~K$. Thus it is clear that the system
goes into a spin glass state with an average freezing temperature,
170~K. Since the particles are coated with gelatin it is clear
that the spin glass behavior can not be due to interparticle
interactions. Rather it has to have its origins within a particle.

 Below $T_{\textnormal{p1}}$, the behavior of this system shows some
features characteristic of superparamagnetic blocking viz.
increase in the FC magnetization on decreasing the temperature,
$H^2$ dependence of $T_{\textnormal{p1}}$ and presence of memory
in FC magnetization without a corresponding effect in ZFC
magnetization. However certain features observed correspond to
spin glass like behavior viz. frequency dependence of
susceptibility with a value of ${\Delta
T_{\textnormal{p}}}/{T_{\textnormal{p}}}$, per decade of frequency
in the range of spin glasses and aging effects in ZFC protocol in
addition to those in FC protocol. Thus the nature of the low
temperature peak is ambiguous.

We have shown that this system shows spin glass behavior in contrast to
most of the earlier reports which claimed superparamagnetism. Further
we have argued convincingly that the reason for this behavior is surface
spin freezing and not interparticle interactions. At low temperature,
below $T_{\textnormal{p1}}$, the behavior shows features of
both superparamagnetism and  spin glasses
thus making its nature  equivocal.

\begin{acknowledgments}
VB thanks the University Grants Commission of India for financial
support.
\end{acknowledgments}

\end{document}